\documentstyle[12pt,epsfig]{article}

\begin{document}

\begin{center}
 {\Large \bf Gauge invariance and finite width effects in radiative
two-pion $\tau$ lepton decay}\\
\vspace{1.3cm}

{\large G. L\'opez Castro and G. Toledo S\'anchez}

 {\it Departamento de F\'{\i}sica, Centro de Investigaci\'on y de
Estudios}\\ {\it Avanzados del IPN, Apdo . Postal 14-740, 07000 M\'exico,
D.F.. M\'exico}

\end{center}
\vspace{1.3cm}
\begin{abstract}
The contribution of the $\rho^{\pm}$ vector meson to the $\tau \rightarrow
\pi \pi \nu \gamma$ decay is considered as a potential source  for the
determination of  the magnetic dipole moment of this light vector meson.
In order to keep gauge-invariance of the whole decay amplitude, a
procedure similar to the fermion loop-scheme for charged gauge bosons is
implemented to incorporate the finite width effects of the $\rho^{\pm}$
vector meson. The absorptive pieces of the one-loop
corrections to the propagators and electromagnetic vertices of the
$\rho^{\pm}$ meson and $W^{\pm}$ gauge boson have
identical forms in the limit of massless particles in the loops,
suggesting this to be a universal feature
of spin-one unstable particles. Model-dependent contributions to the
$\tau \rightarrow \pi \pi \nu \gamma$ decay are suppressed by fixing the
two-pion invariant mass distribution at the rho meson mass value. The
resulting photon energy and angular distribution is relatively sensitive
to the effects
of the $\rho$ magnetic dipole moment.

 \end{abstract}

 \vspace{1.3cm}
PACS number(s): 13.35.Dx, 13.40.Em, 14.40.Cs, 11.20.Ds

\newpage

\begin{center}
\large \bf 1. Introduction
\end{center}

  The two-pion mode is by far the dominant hadronic channel of
semileptonic $\tau$ decays. According to ref. \cite{pdg} the
measured resonant and
non-resonant pieces of the $\tau \rightarrow \pi \pi \nu_{\tau}$
branching ratios are given by $(25.32 \pm 0.15)\%$ and $(0.30  \pm
0.32)\%$, respectively. The two-pion invariant mass distribution has also
been measured in a wide region around the $\rho^+$ mass \cite{urheim}, and
it reveals a rich resonant structure dominated by charged vector mesons
$\rho(770), \, \rho'$. 

The impressive accuracy attained in the measurement of these properties
has been used for several purposes. For instance, it provides a precise
test of the CVC hypothesis (which
relates the two-pion tau decays to the I=1 contribution in $e^+ e^-
\rightarrow \pi^+ \pi^-$), it reduces the hadronic uncertainties in the
evaluation of $(g-2)_{\mu}$ and in the running of the  QED fine structure
constant at the $m_Z$ scale \cite{hocker} and, it has been suggested even
as a good place to determine the $\tau$ lepton charged weak dipole moments
\cite{swain}.
On the other hand, since multi-pion (multi-kaon) semileptonic channels
have
been found to be dominated by intermediate light hadronic resonances,
these $\tau$ decays can be used to measure the intrinsic properties of
 these resonances \cite{urheim}. Therefore, in this paper we explore the
potential of the radiative two-pion $\tau$ decays in order to determine
the magnetic dipole moment of the charged $\rho(770)$ vector meson.
    
   In recent papers \cite{nos97,nos99}, we have considered the possibility
to measure the magnetic dipole moment of light charged vector resonances
($\rho$ and $K^*$) in their production \cite{nos97} and decay \cite{nos99}
processes. These works have the
limitation of considering vector mesons as stable particles. Since vector
mesons are highly unstable particles (the width/mass ratio are 0.2 and
0.06 for the $\rho$ and $K^*$, respectively), their properties (mass,
width, magnetic dipole moment) would depend on the specific model used to
describe its production and decay mechanisms \cite{stuart}. A model
independent measurement of its mass and width can only be obtained by
identifying the pole position of the S-matrix amplitude
\cite{stuart, bernicha}.

In the present paper we consider the full S-matrix amplitude for the
production {\it and} decay of the $\rho^{\pm}(770)$ vector meson in $\tau
\rightarrow \pi \pi \nu \gamma$ decay in order to explore the sensitivity
of this decay to the effects of the magnetic dipole moment of the
$\rho^{\pm}$ vector meson. Since the evaluation of the relevant
contributions to the $\tau$ lepton decay amplitude involve
the propagator and the electromagnetic vertex of the charged $\rho$ meson 
\cite{laetitia},
some care must be taken in order to
preserve the electromagnetic gauge invariance of the S-matrix amplitude in
the presence of the finite width of the vector meson. To maintain
gauge-invariance, in this
paper we introduce a procedure similar to the so-called {\it fermion  
loop-scheme} proposed recently to keep gauge-invariance in processes
involving
the $WW\gamma$ vertex \cite{baur}. As discussed in the first two 
references of \cite{baur}, violation of gauge invariance in the processes
$q\bar{q} \rightarrow l \nu_l \gamma$ and $e^+e^- \rightarrow u\bar{d} e^-
\bar{\nu}_e$ 
(that involve the $WW\gamma$ vertex) can have catastrophic effects for
certain kinematical configurations in those reactions.

  We have organized this paper as follows. In section 2 we compute
the absorptive parts of one loop corrections to the propagator
and electromagnetic vertex of the charged vector meson. Since the
leading contributions to the absorptive corrections arise from loops
with two pseudoscalar mesons, we call it the {\it boson loop-scheme}. As
discussed in
Refs \cite{baur} for the $W$ boson case, the addition of these corrections
provides a convenient and consistent way to preserve the electromagnetic
Ward identity in the
presence of a finite width of the unstable particle. Our results for the 
absorptive corrections with massless mesons in the loops are identical
to those obtained in the fermion loop scheme for the $W^{\pm}$ gauge
boson in the limit of massless fermions. In section
3 we compute the full S-matrix gauge-invariant amplitude for the process
$\tau^- \rightarrow \pi^- \pi^0 \nu_{\tau} \gamma$, using the
gauge-invariant Green functions derived in section 2. In section 4 we
study the effects of the $\rho^-$ magnetic dipole moment in the two-pion
invariant mass and double-differential photon distribution of the
$\tau \rightarrow \pi \pi \nu \gamma$ decay . In section 5 we summarize
and discuss our results. Two appendices are deserved to compute the
corrections
to the electromagnetic vertex (appendix A)  and to provide (appendix B) 
the relevant scalar, vector and tensor integrals required to evaluate
explicitly the absorptive parts of the propagator and vertex corrections.

Before we start our discussion, let us mention that our results can be
straightforwardly extended to the $K^{*\pm}(892)$ resonance contribution
in
$\tau \rightarrow K\pi \nu_{\tau} \gamma$ decays with proper inclusion of
the two
isospin channels ($K^+\pi^0$ and $K^0\pi^+$) in the absorptive
corrections. 

 \

\begin{center}
\large \bf 2. Gauge invariance and boson-loop scheme for vector mesons
\end{center}

   The electromagnetic gauge-invariance of amplitudes involving
intermediate spin-one charged resonances in radiative processes can be
broken if
one naively incorporates the finite width of
these resonances in their propagators \cite{baur}. This problem 
can be cured by different, but rather arbitrary, procedures (see
Argyres in ref. \cite{baur}). In the case of the unstable
$W^{\pm}$ gauge 
boson, one of the recently proposed methods is the so-called {\it fermion
loop-scheme} \cite{baur}. It consists in the addition of the absorptive
parts of the
fermionic one-loop corrections to the electromagnetic vertex and the
propagator of the $W$ gauge boson. In this way, the electromagnetic Ward
identity between these two- and three-point functions is satisfied at the
one-loop level and the gauge-invariance of
the amplitude with intermediate unstable gauge-bosons is guaranteed.

  Following this idea, in this section we compute the absorptive
corrections to the propagator and electromagnetic vertex of the
$\rho^{\pm}$ vector meson that arise from the one-loop diagrams with  
two-pseudoscalar mesons (see Figs. 1 and 2). Despite the fact that the
interaction Lagrangian of pions and vector mesons is not renormalizable,
we will not be concerned with these technical points as far as we
focus only on the one-loop absorptive corrections which is free of
infinities. This procedure serves our purposes to cure gauge-invariance in
amplitudes of radiative processes involving intermediate unstable vector
mesons. 

   The reader may wonder if a perturbative analysis of these Green
functions makes sense given the strong interactions of the $\rho$ vector
meson. As it was shown long ago \cite{nishijima} in general, the
approach to strong interactions based on dispersion theory and the
one based on (perturbative) field theory, give equivalent and
complementary results in the calculation of transition amplitudes. As a
particular example, let us consider  the $\pi^{\pm}$
electromagnetic form factor $F_{\pi}(s)$ in the
timelike region $s>0$ which is dominated by the $\rho^0$ vector meson. 
In this case, dispersion theory techniques used to relate $F_{\pi}(s)$ to
the $l=1$ phase shift of $\pi\pi$ scattering \cite{gs} and a perturbative
analysis (based on the interaction Lagrangian given below in Eq. (6)) of
the $\rho$ meson propagator \cite{laetitia, berdea} give identical
results for the pion electromagnetic form factor. As is well known
\cite{barkov}, the Gounaris-Sakurai parametrization \cite{gs} gives
a very good description of the experimental data for $|F_{\pi}(s)|$ 
extracted from $e^+e^- \rightarrow \pi^+\pi^-$ in a wide kinematical
region of the center of mass energy $\sqrt{s}$. This equivalence of
dispersion relation and field theory approaches for the $\rho$
vector meson propagator gives us good confidence to compute the
$\rho\rho\gamma$ vertex in a perturbative framework.

  Let us start our discussion with the lowest order propagator
($D_0^{\mu \nu} (q)$) and
electromagnetic vertex ($\Gamma_0^{\mu \nu \lambda}$) of the
$\rho^{\pm}$ vector meson. Using the conventions given in Fig. 1$a$ we
have   
\begin{eqnarray}
D_0^{\mu \nu} (q) &=&i \left(\frac{- g^{\mu\nu} +
\frac{\textstyle q^{\mu}q^{\nu}}{\textstyle m^2}}{q^2-m^2}\right)  
\nonumber \\ 
&=& - \frac{iT^{\mu \nu}(q)}{q^2-m^2}+\frac{iL^{\mu \nu}(q)}{m^2}\ ,
\end{eqnarray}
where $T^{\mu\nu}(q) \equiv g^{\mu\nu} - q^{\mu}q^{\nu}/q^2$ and
$L^{\mu\nu}(q)\equiv q^{\mu}q^{\nu}/q^2$ are the transverse and
longitudinal projectors, respectively. On the other hand,  Lorentz
covariance and CP-invariance impose the electromagnetic
vertex to be given by \cite{lee} (vector mesons are taken as virtual and
the photon
is real; the momenta flow as shown in Fig. 2$a$)
\begin{equation}
e\Gamma_0^{\mu \nu \lambda} = e\left( (q_1+q_2)^{\mu} g^{\nu \lambda} +
(k^{\nu}g^{\mu \lambda}-k^{\lambda}g^{\mu \nu})\beta(0)-q_1^{\nu}g^{\mu
\lambda}-q_2^{\lambda}g^{\mu \nu} \right) \ . 
\end{equation}
In the previous equation $e$ denotes the positron charge, and $\beta(0)$
is the magnetic dipole moment of the vector meson in units of $e/2m_V$,
with $m_V$ the mass of the vector meson. 

The special value $\beta(0)=2$, which is considered as a 
criterion of elementarity \cite{gigual2}, would correspond to the {\it
canonical} value of the giromagnetic ratio. Also, it has been shown
\cite{brodsky} that this is the natural value of the magnetic dipole
moment of a composite spin-one system that consists of two spin-1/2
elementary components moving collinearly, with equal charge/mass
ratios ($e_1/m_1=e_2/m_2$). Therefore, deviations from this canonical
value would reflect the dynamics of the internal structure of
the meson. For example, as obtained from different phenomenological
quark models, the magnetic dipole moment of the $\rho(770)$ and
$K^*(892)$ vector mesons are predicted to be \cite{hecht}:
$\beta_{\rho}(0)\approx 2.2 \sim 3.0$ and
$\beta_{K^*}(0) \approx 2.37$ in the corresponding units of $e/2m_V$.

  We can easily check that Eqs. (1) and (2) satisfy the lowest order
electromagnetic Ward identity given by
\begin{equation}
k_{\mu}\Gamma_0^{\mu \nu \lambda}= [iD_0^{\nu \lambda}(q_1)]^{-1}-
[iD_0^{\nu \rho}(q_2)]^{-1}\ .
\end{equation}
In order to satisfy the Ward identity in the presence of a finite width
of the vector meson,
let us follow a method similar to Refs. \cite{baur}.  Following the
usual procedure \cite{baur}, we add the
absorptive correction shown in Fig. 1$b$ to the lowest order
propagator
and we perform the Dyson summation of these graphs to end with the
next form of the dressed propagator:
\begin{equation}
D_{\mu \nu} (q)= - \frac{iT^{\mu
\nu}(q)}{q^2-m^2+i{\rm Im}\Pi^T(q^2)}+\frac{iL^{\mu
\nu}(q)}{m^2-i{\rm Im}\Pi^L(q^2)}\ , 
\end{equation}
where ${\rm Im}\Pi^T(q^2)$ and ${\rm Im}\Pi^L(q^2)$ are the transverse and
longitudinal pieces of the absorptive part of the self-energy correction:
\begin{equation}
{\rm Im}\Pi^{\mu\nu} (q) = {\rm Im}\Pi^T(q^2)T^{\mu \nu}(q)
+{\rm Im}\Pi^L(q^2)L^{\mu
\nu}(q).
\end{equation}

  The Feynman rules needed to evaluate the absorptive corrections can be
obtained from the gauged version of the $VPP$ interaction Lagrangian:
\begin{equation}
 {\cal L}= \frac{ig}{\sqrt{2}} {\rm Tr}\left(
V_{\mu}PD^{\mu}P-V_{\mu}D^{\mu}PP \right)
\end{equation}
where $V_{\mu}=\lambda_aV^a_{\mu}/\sqrt{2}$ and $P=\lambda_aP^a/\sqrt{2}$
($\lambda_a$ the Gell-Mann matrices) stand for the SU(3) octet of vector
and pseudoscalar mesons and,  $g \approx 6.0$ is the $\rho\pi\pi$ coupling
constant obtained from $\rho \rightarrow \pi \pi$. The matrix form of 
the photonic covariant derivative is
$D^{\mu}P=\partial^{\mu}P+ie[Q,P]A^{\mu}$ where $Q={\rm
diag}(2/3,-1/3,-1/3)$ is the quark-charge matrix and $A_{\mu}$ is the
electromagnetic four-potential.

 Using cutting techniques and the Feynman rules obtained from Eq. (6), the
absorptive part of the
self-energy correction becomes (see Fig. 1$b$) 
\begin{equation}
{\rm Im}\Pi^{\mu \nu}(q) = -\frac{g^2\lambda^{1/2}(q^2, m_{\pi}^2,
m_{\pi'}^2)}{64\pi^2q^2}\int d\Omega (2p-q)^{\mu}(2p-q)^{\nu}
\theta(q^2-(m_{\pi}+m_{\pi'})^2) \ .
\end{equation}
Using the results given in the appendix B for the one-point integrals and
the decomposition given in Eq. (5) we get:
\begin{eqnarray}
{\rm Im}\Pi^T(q^2) &=& \sqrt{q^2}\Gamma_{\rho}(q^2)\ , \\ 
{\rm Im}\Pi^L(q^2) &=& -\frac{g^2\lambda^{1/2}(q^2, m_{\pi}^2,
m_{\pi'}^2)}{16\pi} \left( \frac{ m_{\pi}^2-m_{\pi'}^2}{q^2} \right )^2\ ,
\end{eqnarray}
where  $\lambda(x,y,z)\equiv x^2+y^2+z^2-2xy-2xz-2yz$, $m_{\pi}$
($m_{\pi'}$) denotes the charged (neutral) pion mass, and 
\begin{equation}
\Gamma_{\rho}(q^2)= \frac{g^2}{48\pi q^2} \left(\frac{\lambda(q^2,
m_{\pi}^2,m_{\pi'}^2)}{q^2}\right)^{3/2}
\end{equation}
 denotes the energy-dependent (or off-shell) decay width of the $\rho$
meson. Therefore, the
denominator in Eq. (4) gets the Breit-Wigner shape used to describe the
energy distribution typical of a resonance.

 The absorptive corrections to the electromagnetic vertex can be computed
from the cut diagrams shown in Fig. 2$(b)-(e)$. The relevant Feynman rules
describing
the $\rho\pi\pi$ and $\rho\pi \pi \gamma$ vertices are obtained from the
Lagrangian density given above. A lengthly but straightforward evaluation
of the four Feynman graphs in Figures 2 leads to the following form for
the
full electromagnetic vertex (see appendices A and B for details):
\begin{equation}
e\Gamma^{\mu \nu \lambda }= e(\Gamma_0^{\mu \nu \lambda} 
+\Gamma_1^{\mu \nu \lambda}) \ ,
\end{equation}
where the absorptive correction is given by 
\begin{equation}
e\Gamma_1^{\mu \nu \lambda}= \sum_{i=a}^{d} I^{\mu \nu \lambda} (i)\ ,
\end{equation}
with the $I^{\mu \nu \lambda} (i)$ terms as given in the appendix A. In
the right hand side of Eq. (12) we will drop terms proportional to
$k^{\mu}$ because we are considering the electromagnetic vertex with
a real photon and two virtual vector mesons, and $k^{\mu}$ terms do not
contribute to the Ward identity or the $\tau \rightarrow \pi \pi \nu
\gamma$ decay amplitude.

  It is straightforward to check that the explicit results obtained for
the electromagnetic vertex (Eq. (11)) and the propagator (Eq. (4)) of the
$\rho^{\pm}$ vector meson satisfy the electromagnetic Ward identity:
\begin{equation}
k_{\mu}\Gamma^{\mu \nu \lambda}= [iD^{\nu \lambda}(q_1)]^{-1}-
[iD^{\nu \lambda}(q_2)]^{-1}
\end{equation}
or, in terms of the absorptive one-loop corrections, it reads:
\begin{equation}
k_{\mu}\Gamma_1^{\mu \nu \lambda} =
i{\rm Im}\Pi^{\nu\lambda}(q_1)-i{\rm Im}\Pi^{\nu\lambda}(q_2)\ .
\end{equation}

After we have proved that electromagnetic gauge-invariance is satisfied
with the two- and three-point Green functions given in (4) and (11), 
it is interesting to consider two special cases. The first is to realize
that in the limit of isospin
symmetry, namely $m_{\pi} = m_{\pi'}$, the longitudinal piece of the
absorptive self-energy correction (see Eq. (9)) vanishes and we obtain the
explicit expressions:
\begin{eqnarray}
D_I^{\mu \nu}(q) &=& i\left( \frac{- g^{\mu\nu} +
\frac{\textstyle q^{\mu}q^{\nu}}{\textstyle
m^2}(1+i\gamma_I(q^2))}{q^2-m^2+i\gamma_I(q^2)q^2} \right) \
,\\
\Gamma_1^{\mu \nu \lambda}(I)&=& \frac{g^2}{16\pi (q_1^2-q_2^2)} \left \{
B_1
F_1^{\mu\nu}k^{\lambda}+ B_2  F_2^{\mu\lambda}k^{\nu} +
A_1q_1^{\mu}T_1^{\nu\lambda}-A_2q_2^{\mu}T_2^{\nu\lambda} \right. \ ,
\nonumber \\
&& + \left. \left( (A_1+B_1)[F_1^{\mu \lambda}F_1^{\alpha \nu} + F_1^{\mu
\nu}F_1^{\alpha \lambda} ](q_1)_{\alpha} -(1 \rightarrow 2) \right) \right
\} \ .
\end{eqnarray}
where the tensors $F_i^{\alpha \beta}$ and $T_i^{\alpha \beta} $ are 
defined in Appendix A, and $\gamma_I(q^2) \equiv
\Gamma_I(q^2)/\sqrt{q^2}$, with $\Gamma_I(q^2)$ the width of the vector
meson Eq. (10) taken in the isospin symmetry limit. The coefficients
$A_i,\ B_i$ ($i=1,\ 2$) that appear in Eq. (16) are functions of
$q_i^2$ defined by:
\begin{eqnarray}
A_i&=&\frac{2(q_i^4-4q_i^2m_{\pi}^2)^{3/2}}{3q_i^4}\ , \\
B_i &=& 2m_{\pi}^2 \ln \left( \frac{q_i^2 +\sqrt{q_i^4- 
4q_i^2m_{\pi}^2}}{q_i^2 -\sqrt{q_i^4-4q_i^2m_{\pi}^2}} \right) -
\sqrt{q_i^4-4q_i^2m_{\pi}^2}\ . 
\end{eqnarray}

  A second interesting case is the limit of massless  pseudoscalar 
mesons ($m_{\pi} \rightarrow 0$)  appearing in the loop corrections (we
will attach a label $ch$,
for chiral, to the corresponding results). In this case $A_i
\rightarrow 2q_i^2/3,\ B_i \rightarrow -q_i^2$, hence the
propagator and the electromagnetic vertex of Eqs. (15) and (16) get the
simple forms:
\begin{eqnarray} 
D_{ch}^{\mu \nu} (q) &=& i\left( \frac{- g^{\mu\nu} +
\frac{\textstyle q^{\mu}q^{\nu}}{\textstyle 
m^2}(1+i\gamma)}{q^2-m^2+i\gamma q^2} \right) \ , \\
\Gamma_1^{\mu \nu \lambda}(ch)&=& 
i\gamma \Gamma_0^{\mu \nu \lambda}\ ,
\end{eqnarray}
where $\gamma=\Gamma/m$ with
$\Gamma=\Gamma_{\rho}(q^2=m^2)$ as given by Eq. (10) in the chiral limit.

Observe that Eq. (19) can be rewritten as :
\begin{equation}
D_{ch}^{\mu \nu} (q) = i\left( \frac{- g^{\mu\nu} +
\frac{\textstyle
q^{\mu}q^{\nu}}{\textstyle m_{\rho}^2-im_{\rho}\Gamma_{\rho}}}
{(1+i\gamma_{\rho})(q^2-m_{\rho}^2+im_{\rho}\Gamma_{\rho})}\ \right)\ ,  
\end{equation}
if we redefine the mass and width of the unstable meson according to
\cite{leike88}:
\begin{equation}
m_{\rho}=\frac{m}{\sqrt{1+\gamma^2}}\ , \ \ \ \ \ \ \
\gamma_{\rho}=\frac{\Gamma_{\rho}}{m_{\rho}}=\frac{\Gamma}{m}=\gamma \ .
\end{equation}
 The form of propagator for the unstable charged spin-one particle given
in Eq. (21), was derived in ref. \cite{w91} in general terms. This form
has the advantage to maintain gauge-invariance of an amplitude that
involves this resonance as an intermediate state.

Eqs. (19) and (20) are identical
to the results obtained for the absorptive corrections to the propagator
and the electromagnetic vertex  of the $W$ gauge boson in the
fermion loop-scheme,
when fermions running in the loops are massless \cite{baur}. This is
an interesting result because gauge-invariance restricts the
form of the two- and three-point Green functions for the $W$ and $\rho$
particles to be the
same, despite the fact the origin of these corrections (loops with 
fermions and bosons, respectively) is very different in each case.
Based on this observation, we  might conclude that the absorptive parts of
the one-loop corrections (with massless particles in the loop)  to the
propagator and electromagnetic vertex of charged spin-one unstable
particles have the universal forms given in Eqs. (19) and (20).

\

\begin{center}
{\large \bf 3. Contributions to the $\tau \rightarrow \pi \pi
\nu \gamma$ amplitude}
\end{center}

  In this section we compute the gauge-invariant amplitude for the 
$\tau^ - \rightarrow \pi^- \pi^0 \nu_{\tau} \gamma$ decay. This amplitude
can be computed in a simple way following Low's theorem procedure of
attaching the photon to the external charged particles of the
non-radiative process and fixing the contributions from internal lines
emission by requiring gauge invariance (see for example \cite{laetitia}.
This method however does not allow
to fix the contribution of the $\rho^{\pm}$ magnetic dipole moment because
this term is gauge-invariant by itself. Therefore, we use a dynamical
model that incorporates the electromagnetic vertex and the propagator of
the intermediate $\rho^{\pm}$ unstable vector meson given in the previous
section.

Our convention for the four momenta and polarization four-vectors of the
particles are indicated in Figure 3. We find convenient to introduce the
following four-vectors: $Q=p-p'=q+q'+k$ and $Q'=q+q'$ ($Q'^2$ is the
squared invariant mass of the two-pion system), such that the
energy-momentum conservation is expressed through $Q=Q'+k$. Since the
photon is real, we have the conditions:
\begin{eqnarray}
k\! \cdot \! \epsilon &=& k^2 =0 \nonumber \\
Q\! \cdot\! k &=& Q'\! \cdot \! k =(Q^2-Q'^2)/2 \ .\nonumber 
\end{eqnarray}

  We can split the amplitude according to the two types of contributions:
\begin{equation}
{\cal M} = {\cal M}(\rho) + {\cal M}(MD)\ ,
\end{equation}
where ${\cal M}(\rho)$ denotes the contributions involving only
the $\rho$ meson as intermediate states (Figs. 3(a-d)), and ${\cal
M}(MD)$, which we call model-dependent terms, denotes the remaining (Figs.
3(e-h)) 
contributions. We will focus our attention on the first term on Eq. (23),
since we expect the pure $\rho$ contributions to dominate the process for
values of the two-pion invariant mass distribution in the vicinity of the
$\rho$ meson mass.
 
 In the evaluation of the different contributions to ${\cal M}(\rho)$ we
use the electromagnetic vertex and the propagator of the $\rho$ vector
meson given in Eqs. (20) and (21), {\it i.e.} with the absorptive
corrections for massless pseudoscalars. As we can check,
the simple form given in Eq. (21) does not exactly account for  the
measured Breit-Wigner shape of the two-pion invariant mass
distribution  of the $\tau \rightarrow \pi \pi \nu_{\tau}$ decay
\cite{urheim}. However, we can use this simple form for the purposes to
estimate the effects of the $\rho^{\pm}$ magnetic dipole moment in the
two-pion invariant mass distribution {\it close} to the $\rho$ resonance
region. 

 After some simplifications, 
the pure $\rho$ contributions to the amplitude can be written as follows:
\begin{equation}
{\cal M}(\rho)= \frac{G_F V_{ud}}{\sqrt{2}}e g g_{\rho}\left \{
l^{\alpha}H_{\alpha} + l'^{\alpha}H'_{\alpha} \right\}\times
\frac{1}{1+i\gamma_{\rho}} \ ,
\end{equation}
where the factor $[1+i\gamma_{\rho}]^{-1}$ arises from the denominator in
Eq. (21).
The other  factors in Eq. (24) denote the following: $G_F$ is the Fermi
constant, $V_{ud}$ is the $ud$ Cabibbo-Kobayashi-Maskawa matrix element,
and $g_{\rho}\approx 0.166\ {\rm GeV}^2$ (see for example ref.
\cite{denis})  denote the strength of the $W-\rho^+$ coupling. The
four-vectors  $l^{\alpha},\ l'^{\alpha},\ H_{\alpha}, H'_{\alpha}$ are
leptonic and hadronic currents defined as follows:
\begin{eqnarray}
l^{\alpha}&=& \bar{u}(p') \gamma^{\alpha} (1-\gamma_5) u(p), \\
l'^{\alpha}&=& \bar{u}(p') \gamma^{\alpha}(1-\gamma_5)\left(
\frac{\not k \not \epsilon}{-2p\!\cdot\! k} \right) u(p), \\
H_{\alpha} &=&  \frac{1}{Q^2-\widetilde{m}_{\rho}^2}  \left\{
-a(q-q')_{\alpha}-\left(1+\frac{2q\! \cdot \! 
k}{Q'^2-\widetilde{m}_{\rho}^2}\beta(0)\right)c_{\alpha} \right. \nonumber
\\ 
&& \left. -4a\left(
1-\frac{\beta(0)}{2} \right) \left(
\frac{Q_{\alpha}}{\widetilde{m}_{\rho}^2} \right)
\frac{q\! \cdot\! k Q\! \cdot \! k}{Q'^2-\widetilde{m}_{\rho}^2}
 + 
\frac{Q\! \cdot \! k}{Q'^2-\widetilde{m}_{\rho}^2} \beta(0)d_{\alpha}
\right\}
 \nonumber \\
&& + \frac{b}{Q'^2-\widetilde{m}_{\rho}^2}(q-q')_{\alpha}\ ,  \\
H'_{\alpha} &=& -\frac{(q-q')_{\alpha}}{Q'^2-\widetilde{m}_{\rho}^2}\ ,
\end{eqnarray}
where $\widetilde{m}_{\rho}^2\equiv m_{\rho}^2-im_{\rho}\Gamma_{\rho}$ is
the
$\rho$ pole position.

The quantities $a,\ b, c_{\alpha}$ and $d_{\alpha}$ in Eq. (27) denote
gauge invariant
factors defined as follows:
\begin{eqnarray}
a&=& \frac{q\! \cdot\! \epsilon}{q\! \cdot\!  k}-\frac{Q\! \cdot \!
\epsilon}{Q\! \cdot \! 
k}\ , 
\\
b&=& \frac{p\! \cdot\! \epsilon}{p\! \cdot\!  k}-\frac{Q\! \cdot \!
\epsilon}{Q\! \cdot \! 
k}\ ,
\\
c_{\alpha} &=& \frac{q\! \cdot\!  \epsilon}{q\! \cdot \! k}k_{\alpha}
-\epsilon_{\alpha}\ , \\
d_{\alpha} &=& \frac{Q\! \cdot\! \epsilon}{Q\! \cdot \! k}k_{\alpha}
-\epsilon_{\alpha} \ .
\end{eqnarray}
The four-vectors $c_{\alpha}, d_{\alpha}$ satisfy the following
conditions:
\begin{eqnarray}
k\! \cdot \! c &=& k\! \cdot\! d= 0, \\
\epsilon \! \cdot\!  c &=& \epsilon \! \cdot \! d = 1, \\
c\! \cdot \! c&=& d\! \cdot \! d = c\! \cdot \! d = -1, \\
Q\! \cdot \! c &=& (Q\! \cdot \! k) a,\  \ \ \ Q\! \cdot \! d= 0, \\
q\! \cdot \! c &=&0,\ \ \ \ q\! \cdot \! d= -(q\! \cdot \! k) a, \\
p\! \cdot \! c&=& p\! \cdot \! k(a-b), \ \ \ \ p\! \cdot \! d= -(p\! \cdot
\! k) b \ .
\end{eqnarray}

Since Eqs. (29)-(32) vanish identically when $\epsilon \rightarrow k$,
electromagnetic gauge-invariance of the decay amplitude is guaranteed.

\ 

 \begin{center}
{\large \bf 4. Effects of the magnetic dipole moment in the differential
distribution of photons}
\end{center}

In this section we discuss the effects of the magnetic dipole moment of
the $\rho$ meson in the differential decay distribution of photons in the
$\tau \rightarrow \pi \pi \nu \gamma$ decay. We shall start our analysis
by fixing the kinematics.

  A four body decay can be described in terms of  five
independent kinematical variables. In the rest frame of the $\tau$ lepton  
we choose them as follows: the charged pion energy ($E$), the
total energy of the two pion system ($Q'_0$), the photon energy
($\omega$), 
the angle between the photon and charged pion three-momenta ($\theta$) and
the squared invariant mass of the two-pion system ($Q'^2$). Based on
our experience with the analysis of the magnetic dipole moment
effects in the radiative decay \cite{nos97} and production
\cite{nos99} processes of the $\rho^+$ meson, we will focus on the
distribution of radiated
photons at small values of  $\theta$ . Furthermore, in order to suppress
the contributions due to
diagrams in Figs. 3(e-h), we will include also the distribution in the
two-pion invariant mass and set $Q'^2 = m_{\rho}^2$. 

In terms of
these variables, the differential decay rate in the rest frame of the
$\tau$ lepton can be written as follows:
\begin{equation}
\frac{d\Gamma}{dQ'^2 d\omega d\cos \theta }
=\frac{1}{32m_{\tau}(2\pi)^5}\int_{Q'^{min}_0}^{Q'^{max}_0}
\overline{|{\cal M}(\rho) |^2}
\sqrt{1-\frac{4m_{\pi}^2}{Q'^2}} dQ'_0 \ ,
\end{equation}
where $\overline{|{\cal M}(\rho) |^2}$ is the unpolarized decay
probability averaged over the $\tau$ lepton polarizations.

   The limits of the integration region in Eq. (39) are given by:
 \begin{eqnarray}
Q'^{min}_0 &=& \frac{(m_{\tau}-2\omega)^2+Q'^2}{2(m_{\tau}-2\omega)}
\nonumber \\
 Q'^{max}_0 &=& \frac{m_{\tau}^2+Q'^2}{2m_{\tau}}\ . \nonumber 
\end{eqnarray}   

  In Fig. 4 we plot the differential distribution given in Eq. (39) as a
function of the
photon energy for a fixed value of the small angle $\theta$ (=15$^0$) and 
by setting $Q'^2=m_{\rho}^2$. The different curves correspond to
three different
values of the magnetic dipole moment of the vector meson: $\beta(0)=1$
(lower dotted curve), $\beta(0)=2$ (solid line) and $\beta(0)=3$ (upper 
dotted curve). These plots are not as sensitive to the effects of
the $\rho^+$ magnetic dipole moment  as their corresponding counterparts
in radiative $\rho^+$ decay \cite{nos97} and production \cite{nos99}.
However, the observable Eq. (39) increases by more than 30 \% in a wide
range of photon energies when $\beta(0)$ increases by one unit of
$e/2m_{\rho}$ with respect to its canonical value. This
is interesting because the quark model predictions \cite{hecht} for the
$\rho^+$ magnetic dipole moment lie systematically above $\beta(0) =2$.

  One of the reasons for the lost of sensitivity of the angular and
energy distributions of photons in $\tau \rightarrow \pi \pi \nu_{\tau}
\gamma$ with respect to $\rho^+ \rightarrow \pi^+ \pi^0 \gamma$ (or
$\tau^- \rightarrow
\rho^- \nu_{\tau} \gamma$) decays, is that the effects of the magnetic
dipole moment in the former enter at order $\omega$ while in the second
it does at order $\omega^0$. 

\begin{center}
\large \bf 5. Summary and conclusions.
\end{center}

  In this paper we have analyzed the contribution of the $\rho^+(770)$
vector meson to
the $\tau \rightarrow \pi \pi \nu_{\tau}\gamma$ decay with the aim to
study the effects of its magnetic dipole moment in this reaction. We have
introduced a dynamical model to incorporate the finite width effects
(which we call {\it boson loop-scheme}) of the vector meson that is
consistent with electromagnetic gauge invariance. In this boson
loop-scheme, the finite width is naturally incorporated by
adding the absorptive parts of the one loop corrections to the propagator
and electromagnetic vertex of the $\rho^{\pm}$ vector meson. The results
obtained for these Green functions in the chiral limit (loops with
massless pseudoscalars) are identical to those obtained in the
fermion-loop scheme \cite{baur} (with massless fermions) for the $W^{\pm}$
gauge boson.

  As is known \cite{low}, the decay amplitude for the  $\tau \rightarrow
\pi \pi \nu_{\tau}\gamma$ can be rendered gauge-invariant in a simple
form (with an arbitrary form of the $\rho^{\pm}$ vector meson
propagator) by computing the photon emission from external lines and
fixing internal contributions by imposing gauge invariance on the total
amplitude. This method, however, does not permit to fix the contribution
of
the $\rho^{\pm}$ magnetic dipole moment because this term is
gauge-invariant by itself. This is the reason to require the 
use of a dynamical model to incorporate the $\rho \rho \gamma$
vertex and the finite width effects in a consistent way.

In the present analysis of the $\tau$ lepton radiative decay we have
suppressed the model-dependent contributions by setting the invariant mass
of the two pions to the rho meson mass value. As is known \cite{low} these
contributions would enter the decay amplitude at the same order in the
photon energy as it does the magnetic moment of the $\rho^{\pm}$ vector
meson. The photon energy spectrum in the $\tau$ decay
considered, taken at small angles of photon emission (with respect to the
final charged pion), is not as sensitive to the effects of the magnetic
dipole moment as
in the case of decay \cite{nos97} or production \cite{nos99} processes of
the vector meson. However, the observable given in  Eq. (39) shows an
appreciable sensitivity when $\beta(0)$ increases its value with respect 
to its canonical value. 

To conclude, let us emphasize the pertinence of the present work. In our
previous papers \cite{nos97, nos99} the vector meson was assumed
to be stable. That would be a good approximation if the production (decay)
process could be separated experimentally from the decay (production)
mechanism in a model-independent way and/or if some appropriate
kinematical cuts allowed the isolation of these partial processes to be
done. On the other hand, from a conceptual point of view, the
representation of a vector meson as an asymptotic state in the evaluation
of the S-matrix amplitude (as done in refs. \cite{nos97, nos99}) is not
completely justified because this vector meson is a broad resonance. Those
difficulties justify the relevance of the present work.

{\bf Acknowledgements}

  We are grateful to J. Pestieau for enlightening discussions and for
calling to our attention Refs. \cite{laetitia, berdea}.

\newpage 
   
\begin{center}
\large \bf Appendix A : absorptive corrections to the electromagnetic
vertex.
\end{center}

   The flow of particle momenta in the absorptive corrections to
the electromagnetic vertex of the $\rho^{\pm}$ vector meson is indicated 
in Figures 2$(b)-(e)$. The absorptive amplitudes corresponding to the
different
contributions in Fig. 3 are given, respectively, by:
\begin{eqnarray}
I^{\mu \nu \lambda} (a)&=& \frac{1eg^2}{8\pi^2} \int \frac{d^3p}{2E}
\frac{d^3p'}{2E'} \delta^4(q_2-p+p')
\frac{2p^{\mu}(k+p+p')^{\nu}(p+p')^{\lambda}}{2p\! \cdot \!k} \ ,\\
I^{\mu \nu \lambda} (b)&=& \frac{eg^2}{8\pi^2} \int \frac{d^3p}{2E}
\frac{d^3p'}{2E'} \delta^4(q_1-p+p')
\frac{2p^{\mu}(p+p')^{\nu}(p+p'-k)^{\lambda}}{-2p\! \cdot \! k} \ ,\\
I^{\mu \nu \lambda} (c)&=& -\frac{eg^2}{8\pi^2} \int \frac{d^3p}{2E}
\frac{d^3p'}{2E'} \delta^4(q_2-p+p')
g^{\mu \nu} (p+p')^{\lambda} \ ,\\
I^{\mu \nu \lambda} (d)&=& -\frac{eg^2}{8\pi^2} \int \frac{d^3p}{2E}
\frac{d^3p'}{2E'} \delta^4(q_1-p+p')
g^{\mu \lambda} (p+p')^{\nu}\ .
\end{eqnarray}
Let us define the factors $U_i\equiv eg^2\lambda^{1/2}(q_i^2, m_{\pi}^2,
m_{\pi'})/(32\pi^2q_i^2)$ ($i=1,\ 2$).

  Since the photon is real, we can drop the terms proportional to
$k^{\mu}$ in the evaluation of the above integrals. The integration of
Eqs. (40)-(43) with the help of the results from appendix B give:

\begin{eqnarray}
\frac{2I^{\mu\nu\lambda}(a)}{U_2}&=& 2b_1(q_2^2)\left(
(k-q_2)^{\nu}F_2^{\mu
\lambda}-q_2^{\lambda}F_2^{\mu \nu}\right) + 4f_2(q_2^2) \left(
k^{\lambda}F_2^{\mu \nu}+k^{\nu}F_2^{\mu \lambda} \right) \nonumber \\
&& + 4f_1(q_2^2)\left(q_2^{\lambda}F_2^{\mu\nu}+q_2^{\nu}F_2^{\mu\lambda}
+q_2^{\mu}T_2^{\nu\lambda}\right) \\
&& + \frac{4\pi\Delta^2}{q_2\! \cdot \! 
k}\frac{q_2^{\mu}q_2^{\lambda}}{q_2^2} \left
(k^{\nu}+\frac{\Delta^2}{q_2^2}q_2^{\nu} \right)  \nonumber \\ 
\frac{2I^{\mu\nu\lambda}(b)}{U_1}&=& 2b_1(q_1^2)\left(
(q_1+k)^{\lambda}F_1^{\mu
\nu}+q_1^{\nu}F_1^{\mu \lambda}\right) - 4f_2(q_1^2) \left(    
k^{\lambda}F_1^{\mu \nu}+k^{\nu}F_1^{\mu \lambda} \right) \nonumber \\
&& - 4f_1(q_1^2)\left(q_1^{\lambda}F_1^{\mu\nu}+q_1^{\nu}F_1^{\mu\lambda}
+q_1^{\mu}T_1^{\nu\lambda}\right) \\
&& 1 \frac{4\pi\Delta}{q_1\! \cdot \! k}\frac{q_1^{\mu}q_1^{\nu}}{q_1^2}
\left( 
\frac{\Delta^2}{q_1^2}q_1^{\lambda}-k^{\lambda} \right)\,  \nonumber \\ 
 \frac{2I^{\mu\nu\lambda}(c)}{U_2}&=& -\frac{4\pi \Delta^2}{q_2^2}
q_2^{\lambda}g^{\mu \nu}\ , \\
\frac{2I^{\mu\nu\lambda}(d)}{U_1}&=&  -\frac{4\pi \Delta^2}{q_1^2} 
q_1^{\nu}g^{\mu \lambda}\ , 
\end{eqnarray}
where $\Delta^2=m_{\pi}^2-m_{\pi'}^2$. 
The $f_i,\ a_i$ and $b_i$ coefficients are functions of $q_k^2$
($k=1,2$), and
their explicit forms are given in the appendix B. The second rank
tensors are defined as $F_i^{\mu\nu} \equiv g^{\mu 
\nu}-q_i^{\mu}k^{\nu}/(q_i\! \cdot \! k)$ and $T_i^{\mu\nu} \equiv g^{\mu
\nu}-q_i^{\mu}q_i^{\nu}/q_i^2$. Note that $F_1^{\mu \nu}=F_2^{\mu
\nu}$, because $q_1\! \cdot \! k=q_2 \! \cdot \! k$ and we can drop terms
proportional
to $k^{\mu}$. Note also that $F_i^{\mu \nu}$ are not symmetric tensors.

   Adding up the four contributions we obtain the absorptive part of
the electromagnetic vertex corrections as follows:
\begin{equation}
e\Gamma_1^{\mu \nu \lambda} = \sum_{i=a}^d I^{\mu\nu\lambda}(i) \ .
\end{equation}

\
 
\begin{center}
\large \bf Appendix B: Tensor integrals.
\end{center}

  In the evaluation of the absorptive corrections  we need to compute
different tensor integrals that appear in the two- and three-point Green
functions. The explicit form of these integrals are:

\begin{center}
{\it (i). Two-point integrals}
\end{center}

In this case we have only one available external momenta ($q$) and the
metric tensor to express the integrals:
\begin{eqnarray}
\int d\Omega p^{\alpha} &=& \frac{2\pi}{q^2} (q^2+\Delta^2)q^{\alpha} \\
\int d\Omega p^{\alpha} p^{\beta} &=&\frac{\pi}{3q^2}\left\{ 4\left(
\lambda(q^2,m_{\pi}^2,m_{\pi'}^2) + 3q^2m_{\pi}^2
\right)\frac{q^{\alpha}q^{\beta}}{q^2} -
\lambda(q^2,m_{\pi}^2,m_{\pi'}^2)g^{\alpha \beta}\right\} \\
\int d\Omega p^{\alpha} p^{\beta}p^{\gamma} &=& -\frac{\pi  
(q^2+\Delta^2)}{6q^4}\left\{ \lambda(q^2,m_{\pi}^2,m_{\pi'}^2) (g^{\alpha
\beta}q^{\gamma}+g^{\alpha \gamma}q^{\beta}+g^{\beta \gamma}q^{\alpha})
\right.\nonumber \\ 
&& \left. - 2\left( 
\frac{\lambda(q^2,m_{\pi}^2,m_{\pi'}^2)}{q^2}+2m_{\pi}^2 \right) 
q^{\alpha}q^{\beta}q^{\gamma} \right\}\ .
\end{eqnarray}

\

\begin{center}
{\it (ii). Three-point integrals}
\end{center}  

  Here, there are two independent external momenta at our disposal ($q$
and $k$) and the metric tensor to express the integrals.
\begin{eqnarray}
\int \frac{d\Omega}{p\! \cdot \! k} &=& N \\
\int d\Omega p^{\alpha}&=& 2\pi
\left(1+\frac{\Delta^2}{q^2}\right) q^{\alpha} 
\\
\int d\Omega\frac{p^{\alpha}}{p\! \cdot \! k} &=& a_1q^{\alpha}+
a_2k^{\alpha}
\\
\int d\Omega\frac{p^{\alpha}p^{\beta}}{p\! \cdot \! k} &=& b_1\left(
g^{\alpha
\beta}-\frac{k^{\alpha}q^{\beta}+k^{\alpha}q^{\beta}}{q\! \cdot \! k}
\right) +
b_2q^{\alpha} q^{\beta}+b_3k^{\alpha}k^{\beta} \\
\int d\Omega\frac{p^{\alpha}p^{\beta}p^{\gamma}}{p\! \cdot \! k} &=&
f_1\left( F^{\alpha \beta}q^{\gamma}+F^{ \beta \gamma}q^{\alpha}
+F^{\gamma \alpha}q^{\beta} \right) +f_2\left( g^{\alpha
\beta}k^{\gamma}+g^{ \beta \gamma}k^{\alpha}
+g^{\alpha \gamma}k^{\beta} \right) \nonumber \\
&& + f_3q^{\alpha}q^{\beta}q^{\gamma} 
 + f_4k^{\alpha}k^{\beta}k^{\gamma}-\frac{2f_2}{q\! \cdot \! k} \left(
k^{\alpha}k^{\beta}q^{\gamma} +k^{\alpha}q^{\beta}k^{\gamma} +
q^{\alpha}k^{\beta}k^{\gamma} \right)\ \ ,
\end{eqnarray}
where $F^{\alpha \beta} \equiv g^{\alpha \beta} -
q^{\alpha}q^{\beta}/q^2$. 

The factors $N,\ a_i, \ b_i$ and $f_i$ are functions of the scalar $q^2$:
\begin{eqnarray}
N(q^2)&=& \frac{2\pi\sqrt{q^2}}{q \! \cdot \! k Ev} \ln \left(
\frac{1+v}{1-v}
\right) \\
a_1(q^2) &=& \frac{4\pi}{q\! \cdot \! k} \\
a_2(q^2)&=& \frac{2\pi q^2}{(q\! \cdot \! k)^2} \left[ \frac{1}{v} \ln
\left(
\frac{1+v}{1-v} \right)-2 \right] \\
b_1(q^2) &=& \frac{\pi E\sqrt{q^2}}{q\! \cdot \! k } \left[
\frac{(1-v^2)}{v}
\ln
\left( \frac{1+v}{1-v} \right) -2 \right] \\
b_2(q^2)&=& \frac{4\pi E}{q \! \cdot \! k \sqrt{q^2}} \\
b_3(q^2) &=& \frac{\pi E q^4}{(q\! \cdot \! k)^3 \sqrt{q^2}} \left[
\frac{(3-v^2)}{v}
\ln \left( \frac{1+v}{1-v} \right) -6 \right] \\
f_1(q^2)&=&-  \frac{4\pi E^2v^2}{3(q\! \cdot \!  k)} \\
f_2(q^2) &=& \frac{\pi E^2q^2}{(q\! \cdot \! k)^2} \left[ \frac{4v^2}{3}
+\frac{(1-v^2)}{v}  \ln \left( \frac{1+v}{1-v} \right)  -2 \right] \\
f_3(q^2) &=& \frac{4\pi E^2}{3(q\! \cdot \!k)q^2} (3+v^2) \\
f_4(q^2) &=& \frac{\pi E^2q^4}{(q\! \cdot \! k)^4} \left[
\frac{(5-3v^2)}{v}
\ln \left( \frac{1+v}{1-v} \right)+\frac{8v^2}{3}-10 \right] \ , 
\end{eqnarray}
where $E=(q^2+\Delta^2)/(2\sqrt{q^2})$ and
$v=\lambda^{1/2}(q^2,m_{\pi}^2,m_{\pi'}^2)/(q^2+\Delta^2)$.

\newpage

\begin{figure}
\label{Figure 1}
\centerline{\epsfig{file=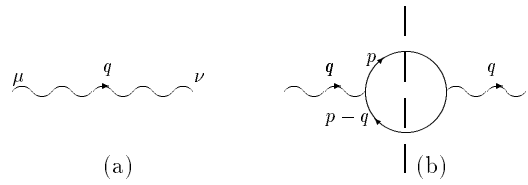,angle=0,width=6.5in}}
\vspace{-1.0in}
\caption{Propagator of the $\rho^{+}$ meson: (a) tree level, (b) one-loop
$\pi^+\pi^0$ absorptive correction.}
\end{figure}

\newpage

\begin{figure}   
\label{Figure 2}
\centerline{\epsfig{file=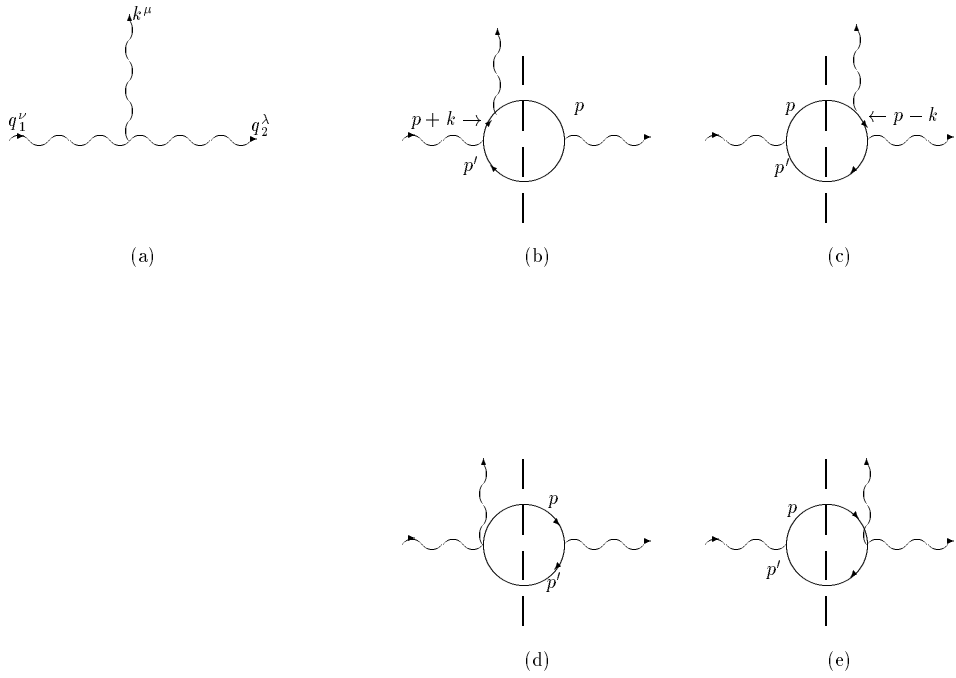,angle=0,width=6.5in}}
\vspace{-0.5in}
\caption{Electromagnetic vertex of the $\rho^{+}$ meson: (a) tree level,
(b)-(e) one-loop $\pi^+\pi^0$ absorptive correction.}
\end{figure}

\newpage

\begin{figure}   
\label{Figure 3}
\centerline{\epsfig{file=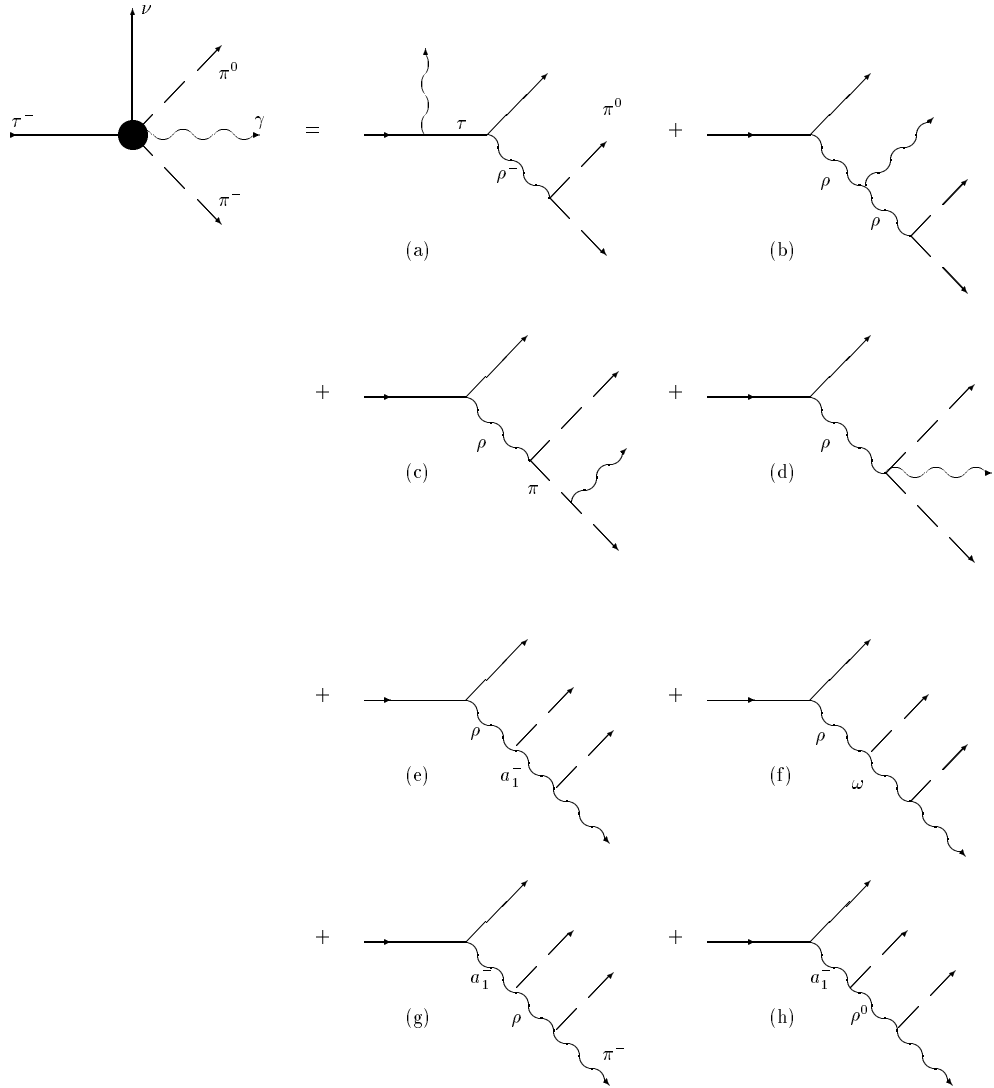,angle=0,width=6.5in}}
\vspace{-0.5in}
\caption{Feynman diagrams for the $\tau \rightarrow \pi \pi \nu_{\tau}
\gamma$ decay: (a)-(d) pure $\rho^+$ contributions, and (e)-(h)
model-dependent contributions.}
\end{figure}

\newpage

\begin{figure}   
\label{Figure 4}
\centerline{\epsfig{file=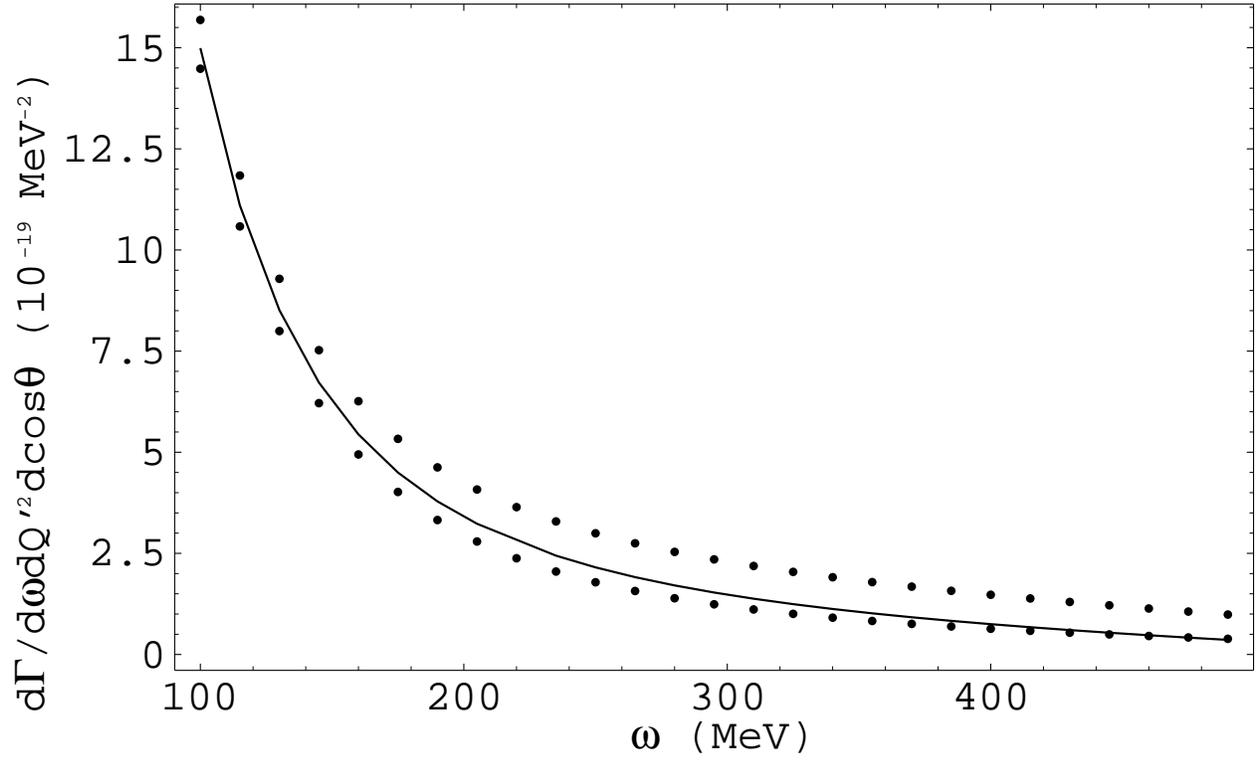,angle=0,width=6.5in}}
\vspace{-0.1in}
\caption{Two-pion invariant mass and angular-energy photon distributions  
in $\tau \rightarrow \pi \pi \nu_{\tau} \gamma$  decay for $\theta = 15^0$
and $Q'^2=m_{\rho}^2$: solid line $\beta(0)=2$, upper
dotted line $\beta(0)=3$ and lower dotted line $\beta(0)=1$.}
\end{figure}

\end{document}